\documentclass{edp-jp4}
\usepackage{psfig,graphicx}
\def\edcomment#1{\iffalse\marginpar{\raggedright\sl#1\/}\else\relax\fi}
\def\b#1{{\bf #1}}\def\d{{\rm d}}
\reversemarginpar
\def\mnras#1 #2 #3 {MNRAS, {\bf#1}, #2 (#3)}
\def\apj#1 #2 #3 {ApJ, {\bf#1}, #2 (#3)}
\def\aj#1 #2 #3 {AJ, {\bf#1}, #2 (#3)}
\def\aap#1 #2 #3 {A\&A, {\bf#1}, #2 (#3)}
\begin{document}

\title{Components of the Milky Way and GAIA} 
\author{James Binney}\address{Oxford University\\
Theoretical Physics, 1 Keble Road, OX1 3NP}
\maketitle
\begin{abstract}
 The GAIA mission will produce an extraordinary database from which we
should be able to deduce not only the Galaxy's current structure, but also
much of its history, and thus cast a powerful light on the way in which
galaxies in general are made up of components, and of how these formed. The
database can be fully exploited only by fitting to it a sophisticated model
of the entire Galaxy. Steady-state models are of fundamental importance even
though the Galaxy cannot be in a steady state. A very elaborate model of the
Galaxy will be required to reproduce the great wealth of detail that GAIA
will reveal. A systematic approach to model-building will be required if
such a model is to be successfully constructed, however. The natural
strategy is to proceed through a series of models of ever increasing
elaborateness, and to be guided in the specification of the next model by
mismatches between the data and the current model.

An approach to the dynamics of systems with steady gravitational potentials
that we call the `torus programme' promises to provide an appropriate
framework within which to carry out the proposed modelling programme. The
basic principles of this approach have been worked out in some detail and
are summarized here. Some extensions will be required before the GAIA
database can be successfully confronted. Other modelling techniques that
might be employed are briefly examined. 

\end{abstract}
%

\section{Introduction}

GAIA will provide at least 5 and often 6 phase-space coordinates for $10^9$
stars. The challenge is to make astrophysical sense of this vast dataset.
Studies of external galaxies have convinced us that galaxies are best
understood as being made up of a series of `components': a bulge or `spheroid' and
perhaps a bar; a thin disk and perhaps a thick disk; a massive halo and
perhaps a metal-poor halo. We must somehow use all those phase-space data to
learn about the components of the Milky Way: how big are they? how old? what
are their radial profiles and shapes?  did they form slowly or suddenly? did
some give rise to others? 

Nearly all galaxies have a disk and a bulge, though the relative importance
of these two components can vary greatly and largely determines the galaxy's
Hubble type. It is probable but not certain that nearly all galaxies have
dark halos. A significant proportion of galaxies possess either a bar or a
thick disk.  Understanding how the different components of galaxies were
formed, and why their relative prominence varies from galaxy to galaxy, are
clearly central questions in the current drive to understand why there are
galaxies at all, and the relationship of galaxies to the rest of the matter
in the Universe.  To answer these questions we need to have the most
complete picture possible of what individual components are, how they the
function dynamically, and how they fit together. The Milky Way, which is a
prototype of the galaxies that are responsible for most of the luminosity in
the Universe, is known to possess all the components listed above, and
kinematic mapping of these with GAIA offers a unique opportunity to clarify
some of the fundamental questions of contemporary astronomy. 

Interpreting the GAIA database in terms of components is a thoroughly
non-trivial task because components are coextensive at many locations, both
in real space and in theoretical spaces in which a kinematic or chemical
datum is used as a coordinate. So it will often be impossible to assign
unambiguously an individual star to this or that component: at best we will
be able to give probabilities for its belonging to one or another component.
Moreover, in the assignment of these probabilities we encounter the
chicken-and-egg problem: until we have assigned stars to
components, we will have a very imperfect knowledge of each component's
chemical composition and dynamics and we will not be in a position to say
how a star's membership probability varies as a function of its chemical and
kinematic data.

For these reasons a major intellectual and computational effort will be
required to pass from the GAIA database to a knowledge of the structure and
dynamics of the Galaxy's components.

\section{The steady-state approximation}

All components are held together by the Galaxy's gravitational potential
$\Phi$, which is currently extremely ill-determined at points away from the
Galactic plane.  A successful attempt to model the various components will
inevitably yield, almost as a spin-off, a good knowledge of $\Phi$
throughout the visible Galaxy.  Taking the Laplacian of $\Phi$ and
subtracting the mass densities of the visible components, we will finally
determine unambiguously the distribution of Galactic dark matter.

The physical principle that will enable us to determine $\Phi$ is that the
Galaxy is in an almost steady state. This assumption, which is only
approximately valid, merits a moment's consideration. We think the Galaxy
should be in an approximately steady state because throughout the visible
Galaxy the dynamical time is orders of magnitude shorter than the Hubble
time, and we have no reason to suppose this is a particularly exciting
moment in the Galaxy's life, such as the climax of a major merger. However,
various processes that are incompatible with a steady state should be
detectable. 

One factor is the bar: deviations from steadiness will be significant unless
we refer everything to the bar's rotating frame. The pattern speed of this
frame is not exactly known, although Dehnen \cite{D99} gives a reasonably
precise value. The bar is almost certainly losing angular momentum to other
components, with the result that it is slowing down and the other components
are heating up. Since bars such as the Galaxy's are extremely common in disk
galaxies, these processes are probably slow and lead to only small
violations of the steady-state principle. The violations are likely to be
detectable, however.

Spiral structure must be redistributing angular momentum within the disk,
and heating it.  This process should lead to small but detectable
violations of the steady-state principle.

The Galaxy is accreting angular momentum that is not aligned with its
current spin axis. This accretion is in the long run expected to lead to
significant reorientation of the spin axis \cite{BM86}, and in the
shorter term probably generates the Galactic warp \cite{JB99},
whose kinematic signature Hipparcos reliably detected for the first time
\cite{D98}.

Finally, the Galaxy is constantly tidally stripping small fry that come too
close and the debris of such stripping will not be in a steady state
\cite{HW99,I00}.

Notwithstanding these process that violate the stead-state approximation,
the latter is a vital tool in the determination of $\Phi$. To see why,
consider the consequence of modelling the GAIA database with a potential that
is much less deep than the true potential. In this case, when the equations
of motion of stars are integrated from the initial conditions that GAIA
provides, the Galaxy will fly apart into intergalactic space. Similarly, if
the adopted potential is too deep, integration of the equations of motion
will result in the Galaxy contracting on a dynamical time, and if the
potential's flattening towards the plane is incorrect, the halo and thick
disk will change shape in the first dynamical time. The true potential is
the one that make the observed stellar distribution pretty much invariant
under integration of the equations of motion.

The idea just described, of integrating the equations of motion forward from
the initial conditions that GAIA will provide, illustrates the physical idea
behind potential estimation, but it is not likely to be useful in practice.
The main reason is that obscuration will prevent GAIA from observing the
entire Galaxy.  Moreover, stars less luminous than the horizontal branch
will not be picked up throughout the Galaxy. If we take the correct
potential and integrate the equations of motion from initial conditions
yielded by such an incomplete survey, the star distribution will {\it not\/}
be invariant: many of the low-luminosity stars that GAIA sees near the Sun
will wander off and will not be replaced because the stars that should
replace them were initially too far away to be seen by GAIA; stars will move
into obscured zones, and gaps in the observed regions will open because
they will not be replaced by stars moving out of obscured zones. Some
more sophisticated procedure is going to be required to test whether the
GAIA data are compatible with the steady-state approximation in a given
potential.

\section{The torus project}

Similar problems arise in accentuated form when one tries to model
ground-based data.  Some years ago my group in Oxford started work on a way of modelling the
Milky Way that promises to overcome these problems \cite{DB96}.
Our work was interrupted by the arrival of the first Hipparcos data and is
only now resuming. It will be disappointing indeed if the project has not
been completed by the time GAIA flies, so I will outline it.

We start from the premise that for any trial potential $\Phi$ we should have
a strictly steady-state dynamical model of each component. We recognize that
real components will not be in exactly steady states, but argue that the
best method of identifying the effects of unsteadiness in the data is
comparison with the best-fitting  steady-state model. Moreover, we hope to be able to model
unsteadiness by perturbing our steady-state model.

Jeans' theorem tells us that a steady-state model of a component may be
generated by taking the component's distribution function (DF) $f_\alpha$ to be
an arbitrary non-negative function of the potential's isolating integrals.
If the potential were `integrable' it would possess just three functionally
independent isolating integrals, and the DF would be a
function of three variables. That is, each component would correspond to a
particular distribution of stars in a three-dimensional space, and the
observed distribution of stars in six-dimensional phase space could in
principle be read off from the function of three variables, just as a living
organism can be constructed from its DNA sequence.

Several practical difficulties have to be overcome before we can exploit
this dramatic simplification. One is that isolating integrals are by no
means unique: a function of two or more isolating integrals is itself an
isolating integral. If we are to talk intelligently about the differences
between the DFs of different components, or the DFs of the same component in
different trial potentials, we must standardize our isolating integrals.
This is readily done by using only action integrals (e.g., \S3.5 of
\cite{BT87}). For an integrable potential these suffer only from a trivial
degree of ambiguity, which is readily eliminated. For an axisymmetric
potential our standard actions are the radial action $J_R$, the latitudinal
action $J_l$ and the azimuthal angular momentum $J_\phi$. The space that has
these actions for Cartesian coordinates we call `action space'. 

In addition to being unambiguously defined for any integrable potential,
actions have the desirable property of faithfully mapping phase space into
action space, in the sense that the volume of phase space occupied by orbits
with actions in the action-space volume $\d^3\b J$ is $(2\pi)^3\d^3\b J$.
Consequently, the DF of a component may be considered to be the density of
stars in action space [up to a factor $(2\pi)^{-3}$] as much as it is the
density of stars in phase space.

Unfortunately, a generic potential will typically not admit three global
isolating integrals, and even if it does, we will not have analytic
expressions for the functions $J_i(\b r,\dot\b r)$ that relate phase space
to action space. Over a number of years my group has developed solutions to
this problem \cite{MB90,BKu93,KB94a,KB94b,K95}. In an integrable potential,
the surfaces in phase space on which actions are constant are topologically
3-tori. On these surfaces the Hamiltonian is constant, and all surface
integrals of the form\footnote{Here $\b p$, $\b q$ are arbitrary canonical
coordinates.} $\int\d\b p.\d\b q$ vanish, so they are called `null-tori'.
Stars move over these null-tori in a rather special way -- each torus admits
three variables, the `angle variables' $\theta_i$, that are canonically
conjugate to the actions that label the tori, and these angles increase
uniformly in time: $\theta_i(t)=\theta_i(0)+\omega_i t$.  Unless the
frequencies $\omega_i$ are commensurable, it follows that in a steady-state
model the phase-space density of stars is constant over a torus: this is the
origin of Jeans' result that the DF of a steady-state model does not depend
on the $\theta_i$.

It turns out that if all phase space can be foliated by such null 3-tori,
and the given Hamiltonian $H$ is constant on them, then $H$ is integrable,
and its actions label the tori by giving the magnitudes of their three
independent cross-sectional areas, $J_i=(2\pi)^{-1}\oint_i\b p.\d\b q$. We
have developed a technique for foliating phase space with null tori on which
a given $H$ is nearly constant.  These tori can be used to define a
integrable Hamiltonian $\hat H$ that differs from $H$ by only a small amount.

\subsection{Resonances \& perturbation theory}

A real galactic potential is exceedingly unlikely to be integrable in the
sense that it admits a global set of angle-action variables. Consequently,
the integrable Hamiltonian $\hat H$ will surely differ from the true
Hamiltonian $H$ at some level.  Since $\delta H=H-\hat H$ is small compared
to $H$, stars integrated in $H$ and $\hat H$ from the same initial condition
will stay close to one another for a few orbital times. In fact, the motion
in the true Hamiltonian $H$ can be considered to be the result of perturbing
the integrable Hamiltonian $\hat H$ by $\delta H=H-\hat H$. The astronomical
significance of this perturbation will depend on the age of the system
measured in dynamical times and whether orbits of interest lie near a
resonance of $\hat H$: if the initial conditions lie on a resonant torus,
even the small perturbation $\delta H$ can cause the orbit in $H$ to deviate
significantly from that in $\hat H$ if you wait long enough. Consequently,
the phenomena occurring in $H$ can differ significantly from those occurring
in $\hat H$, and it may be necessary to obtain a better approximation to the
true dynamics than $\hat H$ provides. 

\begin{figure}
\centerline{\psfig{file=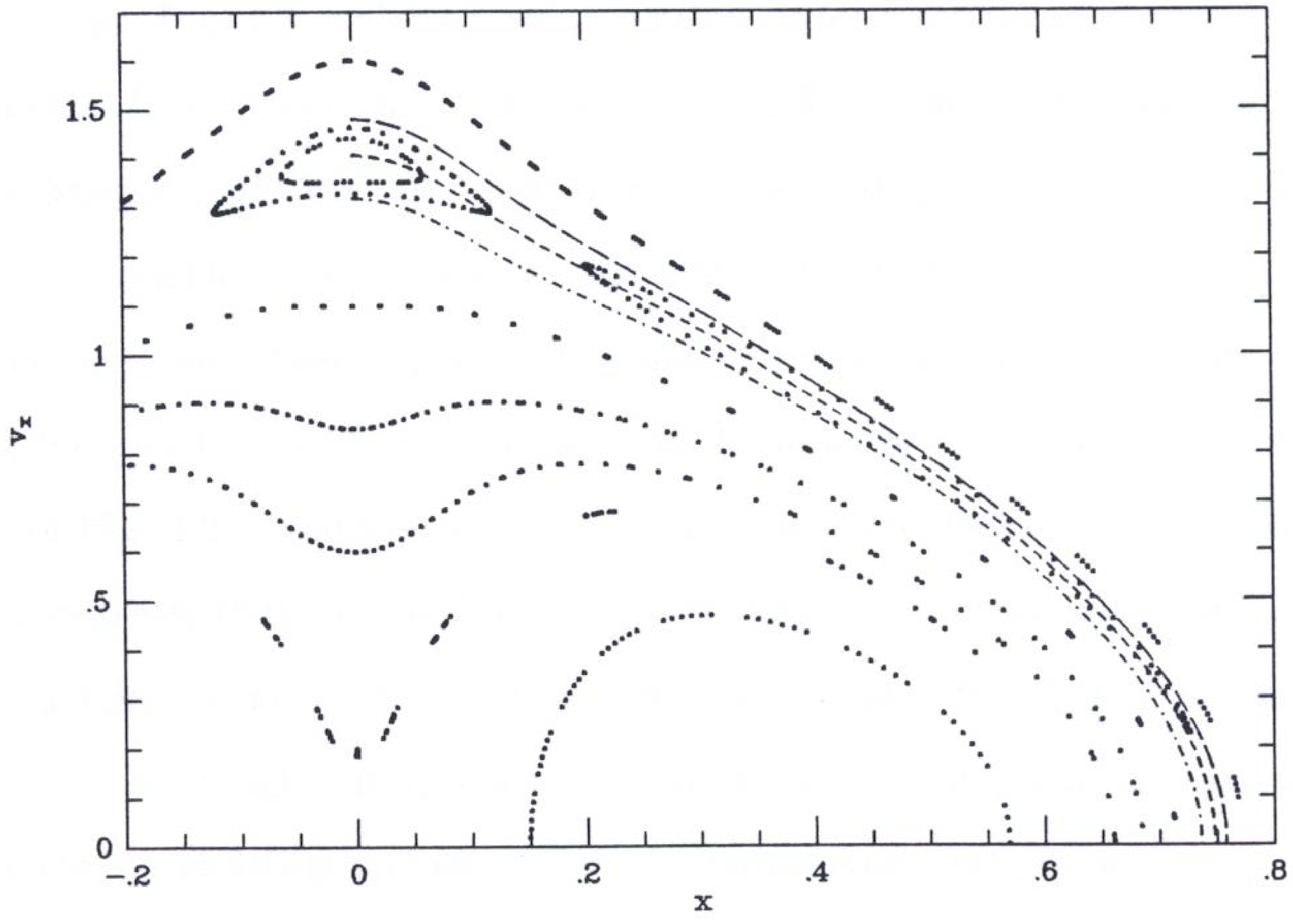,width=.48\hsize}\hfill
\psfig{file=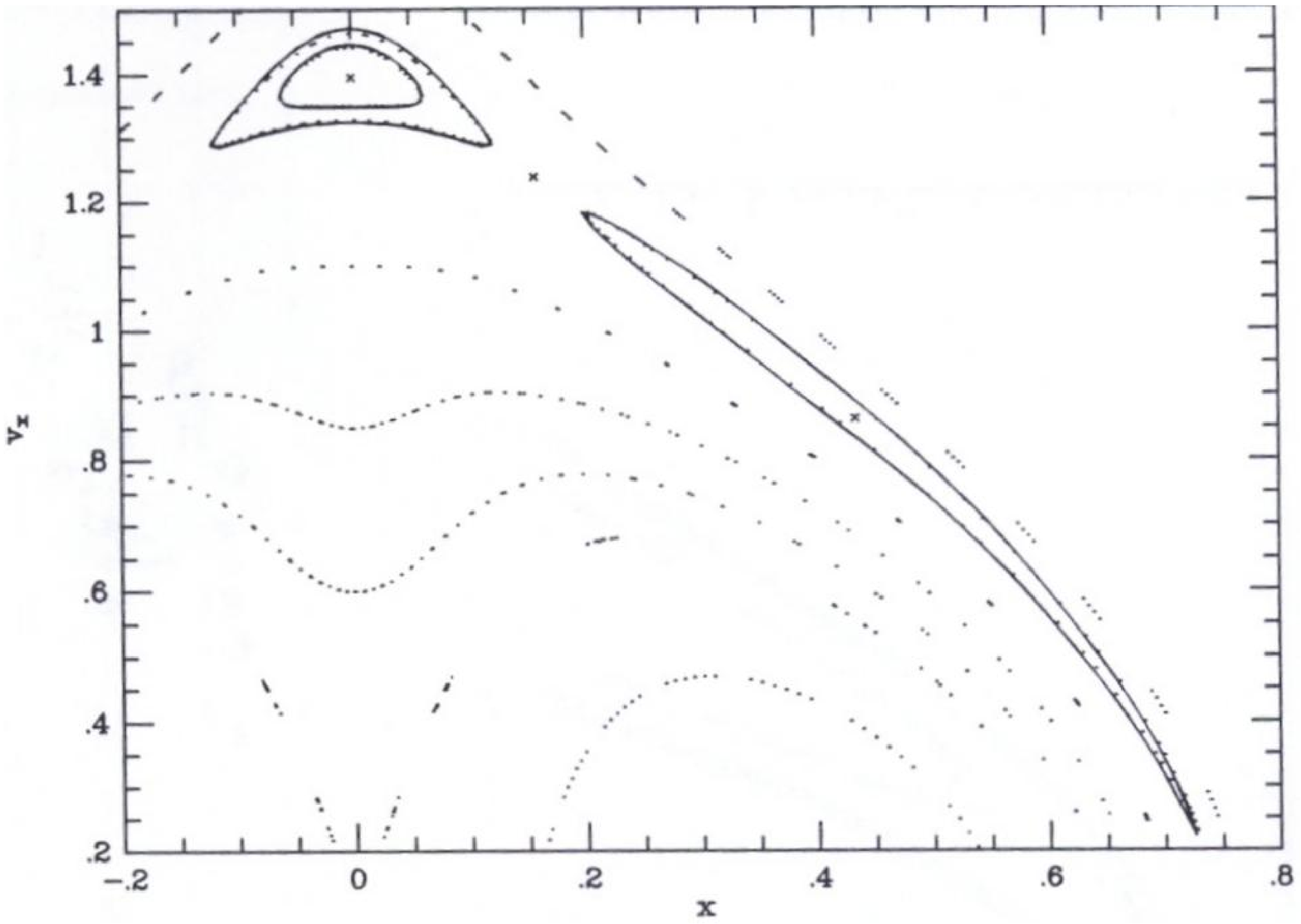,width=.48\hsize}}
\caption{A surface of section for motion in a barred potential. Dots show
consequents of numerically integrated orbits. In a band towards the outside
of the figure, these delineate a chain of islands. At left three dashed curves
through these islands show cross sections of tori of the underlying
integrable Hamiltonian $\hat H$. At right the full curves that accurately
delineate the islands are obtained by treating $\delta H$ as a perturbation
on $\hat H$. (From \cite{Kthesis})\label{Kfig}}
\end{figure}

\begin{figure}
\centerline{\psfig{file=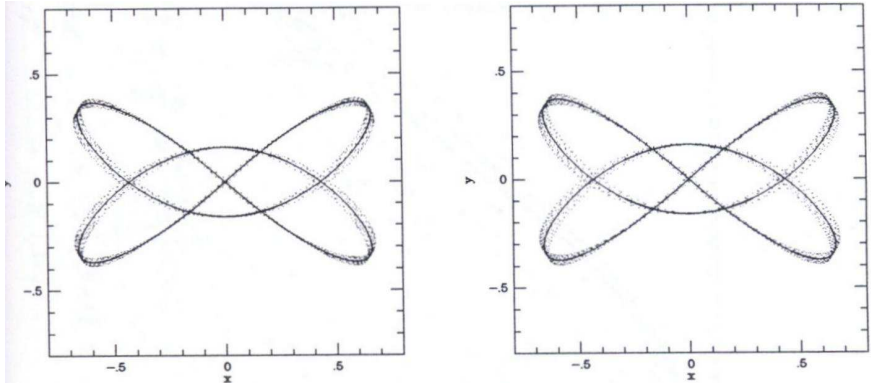,width=.8\hsize}}
\caption{One of the 2:3 resonant box orbits that generate the chain of
islands in Figure \ref{Kfig}. The orbit shown at left was generated by
direct integration, while that at right was generated by applying
perturbation theory to the integrable Hamiltonian $\hat H$. The full curve
is the underlying closed orbit. (From \cite{Kthesis})\label{Korbits}} 
\end{figure}

The torus programme offers two ways of improving on the model provided by
$\hat H$. The left panel of Figure \ref{Kfig} illustrates one method by
showing part of a surface of section. The figure's dots are the consequents
of eight orbits in a barred gravitational potential. These orbits all admit
an isolating integral in addition to $H$ because their consequents lie on
smooth curves.  Six of these curves have the characteristic shapes of the
invariant curves of box and loop orbits (e.g., Fig.~3-8 in \cite{BT87}),
which are associated with a global system of action-angle coordinates (\S3-5
of \cite{BT87}).  Just inside the outer-most invariant curve, the invariant
curves have a different structure, forming part of what would be a chain of
six islands if the whole surface of section were plotted. The torus
machinery has been used to draw three dashed curves through the region
occupied by the islands: one curve goes through the middle of the islands,
while flanking curves pass either side of them. These curves are invariant
curves of $\hat H$, which admits global action-angle coordinates and
therefore supports only boxes and loops.

The right panel of Figure \ref{Kfig} shows the same surface of section, but
now with the islands delineated by full curves. These curves are generated
by treating $\delta H$ as a perturbation of $\hat H$. Since $\delta H/H$ is
small, the agreement between the numerical consequents and the invariant
curves from first-order perturbation theory is excellent.\footnote{To
exploit fully the smallness of $\delta H/H$, Kaasalainen \cite{Kthesis} had
to develop an extension of standard first-order perturbation theory.} Figure
\ref{Korbits} shows that in real space one cannot distinguish between the
orbits obtained by direct integration and perturbation theory.

\begin{figure}
\centerline{\psfig{file=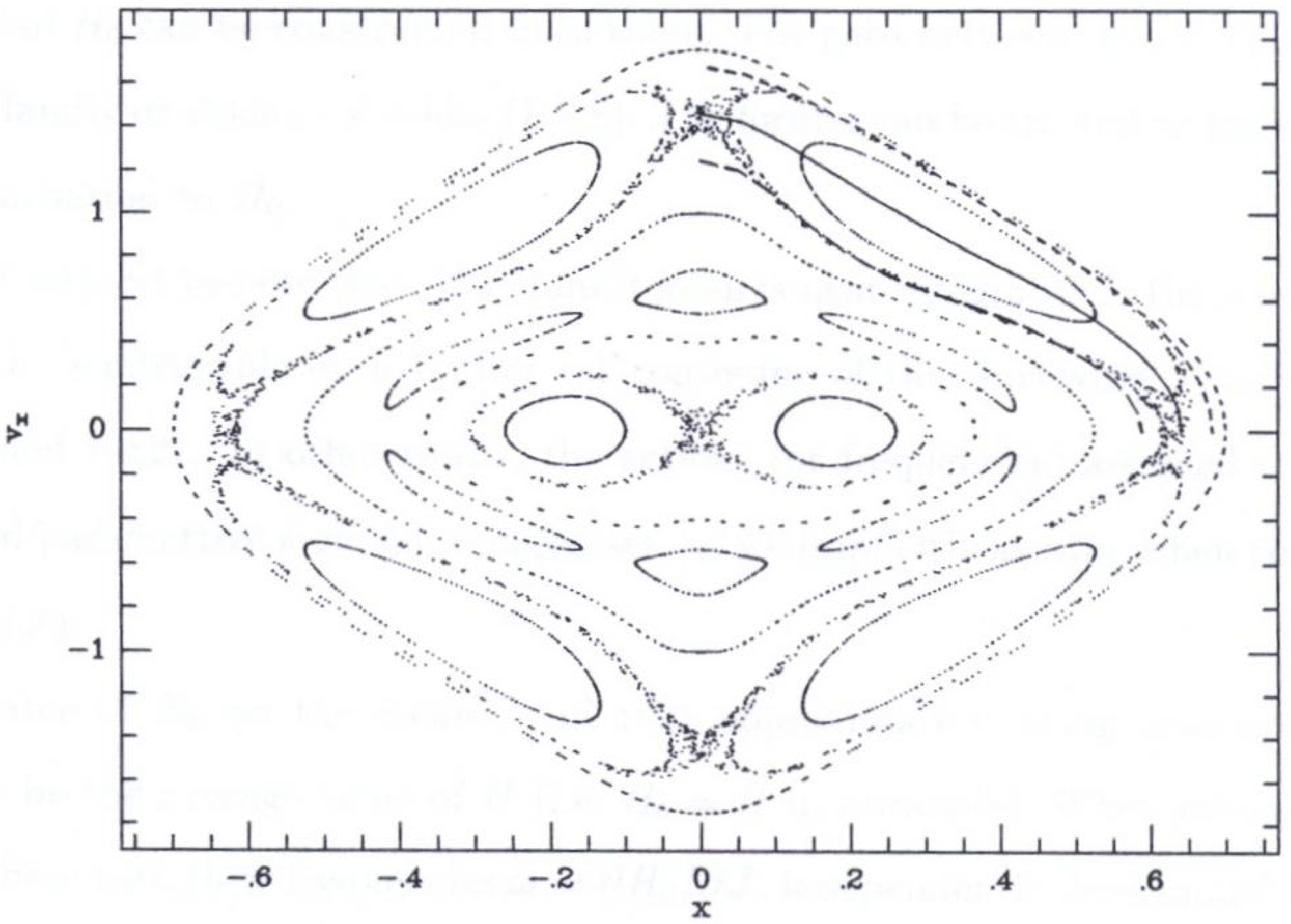,width=.48\hsize}\hfill
\psfig{file=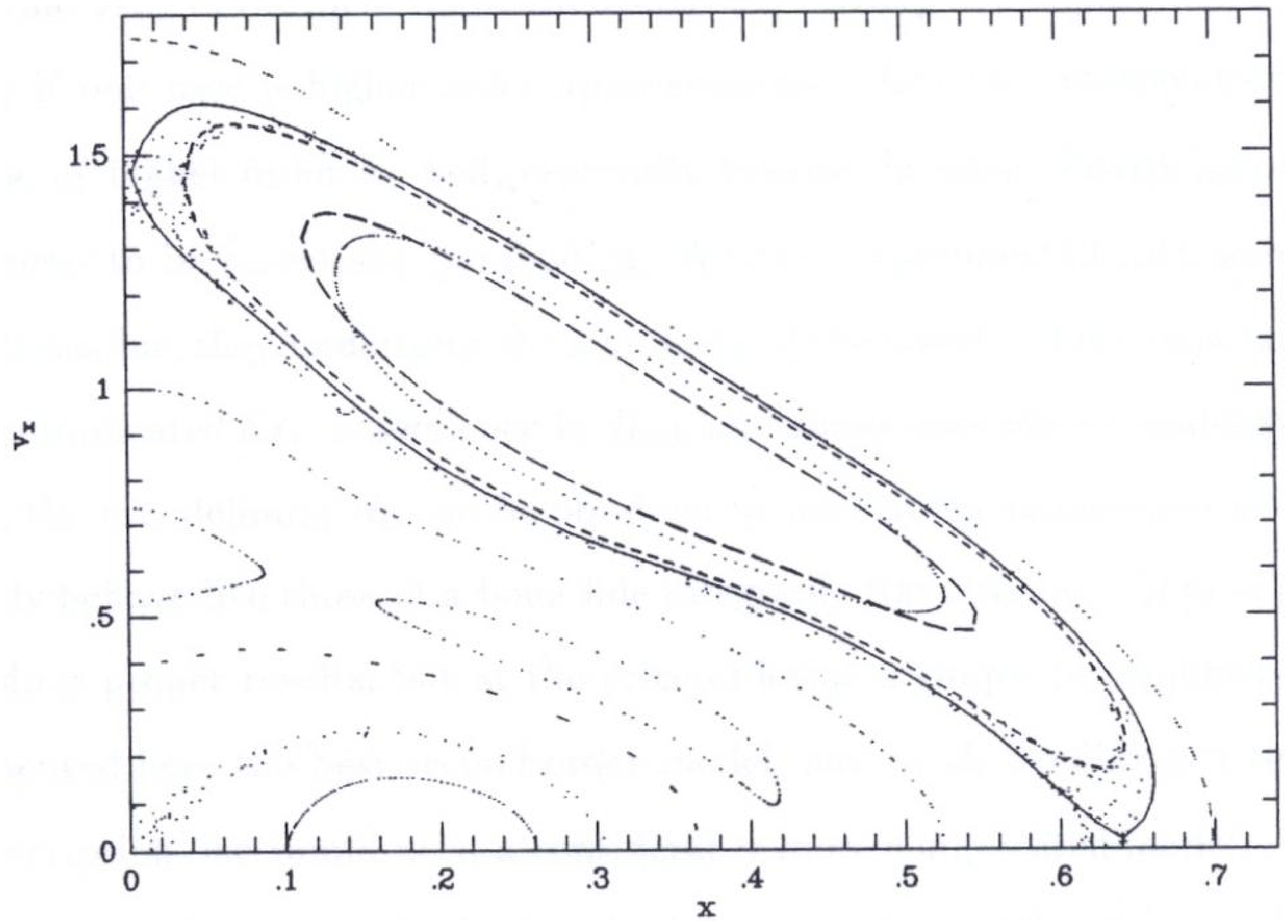,width=.48\hsize}}
\caption{A surface of section in a barred potential in which most orbits are
resonant boxes (`boxlets'). The curves in the left panel shows the tori
generated if all orbits are assumed to be non-resonant boxes. The curves in
the right panel shows the tori obtained when orbits are assumed to be
resonant. (From \cite{Kthesis})\label{Kfig3}}
\end{figure}

Figure \ref{Kfig3} shows an alternative approach to resonant orbits, which is
appropriate when the resonance is powerful, and therefore $\delta H/H$ is
not small. The left panel again shows invariant curves of $\hat H$ ploughing
through the resonant region, which is now associated with non-negligible
stochasticity near the seperatrices. The right panel shows excellent
agreement between consequents for a series of resonant orbits and invariant
curves obtained by generating a new integral hamiltonian $\hat H'$ for the
resonant region. This Hamiltonian is obtained by assuming that tori have the
general structure required for libration about the closed resonant orbit,
and then deforming them so as to make $H$ as nearly constant over them as
possible. 

These examples show that the torus programme can, in principle, provide an
extremely accurate description of regular orbits, no matter what their
structure. The ability of the programme to give a good account of stochastic
orbits has not been so completely explored. I anticipate, however, that it
will prove powerful in this area also, since, as Figure \ref{Kfig3}
illustrates, it can provide a system of action-angle variables in which the
stochastic region is bounded by particular tori. Let us call the values
taken by $\b J$ on these tori the `critical actions'.  It is likely that
over time the DF will tend to a function of energy only
in the region of action space that is bounded by the critical actions.

\subsection{Galaxy modelling with tori}

Our procedure for interpreting a data set such as GAIA will produce is as
follows. We start with a trial potential $\Phi(\b r)$. We foliate phase
space with 3-tori on which $H=\frac12 {\dot r}^2+\Phi(\b r)$ is nearly
constant. This foliation establishes a system of action-angle coordinates
for an integrable Hamiltonian $\hat H$ that differs from $H$ only slightly.
Routines are then available to pass between these action-angle coordinates
and the ordinary Cartesian coordinates for phase space,  $(\b r,\dot\b r)$.

Next, for each component $\alpha$ of the Galaxy we choose a simple
analytic DF $f_\alpha(\b J)$. Because the actions are
unique and physically well-motivated variables, it is easy to understand the
relationship between the form of $f_\alpha$ and the observables of the
component, such as its flattening, characteristic spin and the typical
eccentricity of its stellar orbits \cite{BCamb}. Finally, for each component
we choose an Initial Mass Function and a star-formation history, which
together enable us to predict the distribution in colour and absolute
magnitude of the component's stars.

We now have a steady-state dynamical model of each component in the given
$\Phi$. This model predicts the probability of observing a star of a given
component at any phase space location. By convolving this probability
distribution with the assumed colour and absolute-magnitude distributions of
the component, and summing over components, we convert these probabilities
into the probability of finding a star of given colour and $M_V$ at any
point in phase space.  Finally, these probabilities are convolved with the
selection functions in colour and phase space of any given survey. This
final probability distribution is then evaluated at the location of each
catalogued star and the resulting numbers are multiplied together to give
the likelihood of the catalogue given the current Galaxy model.  We propose
to maximize this likelihood by adjusting a suitably parameterized form of
the Galactic potential $\Phi(\b r)$ as well as the functions that
characterize each component: the function of three variables $f_\alpha(\b
J)$, the IMF and the star-formation history. This maximization is likely to
be a computationally challenging task, but not one that is out of proportion
to the other computational challenges that GAIA inherently poses.
Notice that the final model will
encode not only the current state of the Galaxy, but much information about
its past. Some more detail and sample calculations can be found in
\cite{DB96}.

\section{Other modelling techniques}

It is not self-evident that the approach just described to modelling the
GAIA database will be the most important one in practice, but  it
does contain a number of elements that {\it any\/} viable
technique is likely to include.

First, I believe it is essential to produce a steady-state model of the
Galaxy. Such a model is an unrealized ideal, but a key step both in the
determination of the Galactic potential, and in deducing what features in
the data are symptomatic of unsteady dynamics. 

Second, one has somehow to extrapolate the stellar distribution from the
parts of the Galaxy that are observed, to those that are not. The strategy
suggested above for doing this is three-fold. First Jeans' theorem is used
to argue that if I observe the stellar density at one point in phase space,
I know the density at all other points in phase space at which the isolating
integrals take the same values. For hotter components, such as the halo and
the thick disk, this argument allows one to determine even from observations
confined to the solar neighbourhood, the value of the DF through a
surprisingly large part of phase space \cite{MB86}. For more luminous stars,
GAIA's coverage will be so extensive that this principle will be very
powerful indeed. Second, the theory of stellar evolution and nucleosynthesis
is used to connect the phase-space densities of faint stars to those of
their more luminous brethren. Finally, the uniqueness of the action
integrals is used to reduce the DF of each component to an analytic function
of the actions that depends on a small number of parameters. This reduction
not only simplifies the computational task of optimizing the DFs, but also
facilitates astrophysical interpretation of the results.

\subsection{Schwarzschild's modelling technique}

A widely used technique for modelling external galaxies is that of
Schwarzschild \cite{Schw79,Schw81} and it is interesting to  examine the
possibility of using this to model the GAIA database. In Schwarzschild's
method one again starts with a trial potential $\Phi$, but one integrates a
large number of orbits in it instead of constructing tori for it. Then,
instead of choosing a DF $f_\alpha$ for each component, one chooses a set of
weights $w^{(i)}_\alpha$ for each orbit $i$. The merit of the method is that
orbit integration is computationally simple, and uses routines that are
completely independent of the nature of the orbits: whether they are tube or
box orbits, regular or stochastic.  Torus construction, by contrast, has to
be tuned to the dominant orbit families. Moreover, Schwarzschild's method
deals properly with families of resonant orbits whereas torus construction
sweeps these under the rug for possible subsequent examination by
perturbation theory.

Schwarzschild's method has several weaknesses, however. One is that it is
cumbersome numerically because phase-space points have to be stored at many
points along each orbit, with the result that an `orbit library' of 10,000
orbits will occupy of order $1\,$Gb on disk.  Moreover, the resolution in space and
velocity of the final model is determined by the number of orbits and the
temporal frequency at which each is sampled. In the torus method, by contrast,
each torus is represented by a relatively small number of expansion
coefficients from which phase-space points can be evaluated dynamically as
the model is compared to the data.  There is no limit to how densely a given
torus is sampled, and once a reasonable torus library is to hand, additional
tori can be quickly constructed without reference to the Hamiltonian by
interpolation on the expansion coefficients for nearby tori in the library.
Finally,  in its classical form Schwarzschild's method gives no insight into
which orbits are `close' to each other in phase space. This has two
consequences. One is that there is no way of requiring the weights of
neighbouring orbits to be nearly equal, as seems physically reasonable. The
other is that one cannot determine the density of a component at a given
point in phase space, which makes it impossible to  compare the phase-space
structure of  models built with different orbit libraries, even if the
potentials are identical. In fact, communication of a model requires
transmission of both the $\sim1\,$Gb of the orbit library and a complete set
of orbit weights ($\sim100\,$Kb per component). By contrast, a model
constructed by the torus method can be communicated by tabulating the four
or five parameters in the DF of each component.

H\"afner et al \cite{H00} show how Schwarzschild's method may be upgraded to the
point at which it returns the DF at the location of each orbit.  Moreover
following Zhao \cite{Z96}, one can assign `effective integrals' to each orbit
which enable one to say, in an approximate way, which orbits are close to
one another, and thus impose continuity of the DF. Moreover, one could insist
that the weights were those implied by an analytic function of the effective
integrals that depends on a small number of parameters, in the same way
that the torus method assumes the DF to be a parameterized function of the
actions. Used in this mode, Schwarzschild's method could be equal to the
task of modelling the GAIA data set.

\subsection{N-body models}

Fux \cite{F99} has made a significant contribution to our understanding of the
dynamics of the inner Galaxy simply by observing a suitable $N$-body model.
Could this approach make a significant contribution to our understanding of
the GAIA dataset? If we were to set up an $N$-body model simply by using the
coordinates returned by GAIA as initial conditions, we would run up against
the problems with observational selection that were described above. A
better way of choosing initial conditions would be to start by fitting to
the GAIA data to DFs of the form $f_\alpha(\b r,\dot\b r)=\rho_\alpha(\b
r)p_\alpha(\dot\b r)$, where $\rho_\alpha$ is an analytic fit to the density
distribution of component $\alpha$ and $p_\alpha$ is an analytic probability
density that approximately describes the distribution of velocities within
this component. For judiciously chosen $\rho_\alpha$ and $p_\alpha$ the
initial conditions might soon settle to a steady-state that resembled the
Galaxy. Setting up an $N$-body model in this way would be by no
means trivial, however, because choosing the functions $p_\alpha(\dot \b r)$ is likely
to be a delicate business. The technique
devised by Syer \& Tremaine \cite{ST96}, in which the masses of particles are
dynamically adjusted, may be able to make up for short-comings in the choice
of the $p_\alpha$. 

All of these particle-based schemes -- Schwarzschild's technique, and N-body
modelling with or without the refinements of Syer \& Tremaine -- will suffer
from the drawback that, for feasible numbers of masses in the model, errors
in the model's observables, such as velocity distributions near the Sun,
will far exceed those in the data.  Consequently, none of these schemes is
likely to do justice to the precision of the GAIA data. 

\section{Conclusion}

GAIA poses an enormous challenge to the theorist because it is essential
that its vast data set be modelled as a whole and in a single sweep. The
modelling must include not only the dynamics of the contemporary Galaxy, but
many aspects of its history as well, most particularly the star-formation
history of its various components.

In view of the scale of this enterprise, it is fortunate that the data will
not arrive for more than a decade. In that period Moore's law for the growth
of computer power will ensure that the necessary computational resources
will be available. If we start now, there is a reasonable chance that
appropriate computational schemes will also be on hand for modelling the
Galaxy in the requisite depth and breadth. Developing these schemes will be
astronomically rewarding in the short term also, since there is an abundance
of ground-based data to model that poses the same conceptual problems in
heightened form.

The richness of the GAIA database will ultimately allow us to study the
Galaxy in exquisite detail, and to learn about the various chance events
that have cumulatively shaped it. Extracting this detail from the database
will require subtlety, however, and it is likely the that best strategy for
mining the database will be one in which models of systematically increasing
sophistication are successively fitted to the data. The first models would
assume that the Galaxy has a globally integrable potential and is in a
strictly steady state. Discrepancies between the data and the best-fitting
model of this type might indicate that certain resonances are not to be
ignored. At the next level a model that included these resonances but was
still in a strictly steady-state would be fitted to the data.
Discrepancies between model and data might now point to non-steady phenomena
such as spiral structure. Perturbation theory would then be used to model
these effects, and discrepancies again sought. Proceeding in this manner one
can imagining constructing a very detailed model that reflected many of the
chance events that have fashioned the Galaxy, as well as ongoing evolution
driven by the bar, spiral structure and the Sagittarius dwarf galaxy.

The torus programme has a number of features that suit it very well to this
programme of work. Most importantly, it allows one to start with an
extremely simple model that can be described by only a handful of
parameters, and to upgrade this model through a systematic sequence of well
defined stages. At each stage, the model fitted to the data at the
preceding stage provides a clear basis from which to advance to a more
elaborate model. Another important advantage of the torus programme is the
facility to beat discreteness noise down to any predefined value in a
straightforward way.

A number of published papers demonstrate the basic principles of the torus
programme for the case of two-dimensional potentials, which effectively
includes all three-dimensional axisymmetric potentials. The generalization
of these principles to general, nearly integrable, three-dimensional
potentials is straightforward though computationally costly. Important tasks
that must be accomplished before the torus programme can be applied to the
GAIA database include exploring its application to chaotic orbits and,
through perturbation theory, to non-steady systems.


\end{document}